\input harvmac
\input epsf
\def\ev#1{\langle #1 \rangle}
\def\co{\Lambda_0}

\def\N{{\cal N}}

\def\tym{\tau_{\rm YM (\co )}}
\def\F{{\cal F}}
\def\a{\alpha}
\def\b{\beta}
\def\g{\gamma}

\Title
 {\vbox{
 \baselineskip12pt
 \hbox{HUTP-02/A021}
 \hbox{hep-th/0206017}\hbox{}\hbox{}
}}
 {\vbox{
 \centerline{$\N=1$ and $\N=2$ Geometry from Fluxes}
 }}
 \medskip

\centerline{F. Cachazo and
C. Vafa}
\medskip
\centerline{Jefferson Physical Laboratory}
\centerline{Harvard University}
\centerline{Cambridge, MA 02138, USA}
\bigskip

\vskip .3in \centerline{\bf Abstract}
We provide a proof of the equivalence
of $\N=1$ dynamics obtained by deforming $\N=2$
supersymmetric gauge theories by addition of
certain superpotential terms, with that of type IIB 
superstring on Calabi-Yau threefold
geometries with fluxes.  In particular we show
that minimization of the superpotential involving
gaugino fields is equivalent to finding loci where
Seiberg-Witten curve has certain factorization property.
Moreover, by considering the limit
of turning off of the superpotential we obtain the full
low energy dynamics of $\N=2$ gauge systems from Calabi-Yau
geometries with fluxes.
 \smallskip
\Date{June 2002}

%%%%%%%%%%%%%%%%%%%%%%%%%%%%%%%%%%%%%
\lref\tva{T.R. Taylor and C. Vafa, ``RR Flux on Calabi-Yau and
Partial Supersymmetry Breaking'', hep-th/9912152,
Phys.Lett. {\bf B474} (2000) 130-137}
\lref\mayr{P. Mayr,``On Supersymmetry Breaking in String Theory and its
Realization in Brane Worlds'',
hep-th/0003198, Nucl.Phys. {\bf B593} (2001) 99-126}
\lref\swc{N. Seiberg and E. Witten,``Monopole Condensation, And Confinement
In N=2 Supersymmetric Yang-Mills Theory'', hep-th/9407087, Nucl.Phys. {\bf
B426} (1994) 19-52;
P.C. Argyres and A.E. Faraggi,``The Vacuum Structure and
Spectrum of N=2 Supersymmetric SU(N) Gauge Theory'', hep-th/9411057,
Phys.Rev.Lett. {\bf 74} (1995) 3931-3934;  A. Klemm, W. Lerche, S. Theisen,
and S. Yankielowicz, ``Simple Singularities and N=2 Supersymmetric
Yang-Mills Theory'', hep-th/9411048, Phys.Lett. {\bf B344} (1995) 169-175}
\lref\cfkiv{ F. Cachazo, B. Fiol, K. Intriligator, S. Katz, and C. Vafa,
``A Geometric Unification of Dualities'', hep-th/0110028,
Nucl.Phys. {\bf B628} (2002) 3-78}
\lref\civ{F. Cachazo, K. Intriligator, and C. Vafa, ``A Large N Duality via
a Geometric Transition'', hep-th/0103067, Nucl.Phys. {\bf B603} (2001) 3-41.}
\lref\ckv{F. Cachazo, S. Katz, and C. Vafa, ``Geometric Transitions and
N=1 Quiver Theories'', hep-th/0108120}
\lref\vafa{C. Vafa, ``Superstrings and Topological Strings at Large N'',
hep-th/0008142, J. Math. Phys. {\bf 42} (2001) 2798-2817}
\lref\gvw{S. Gukov, C. Vafa, and E. Witten, ``CFT's From Calabi-Yau
Four-folds'', hep-th/9906070, Nucl.Phys. {\bf B584} (2000) 69-108}
\lref\kra{See for example: H. Farkas and I. Kra, ``{\it Riemann Surfaces}''
(Springer-Verlag, 1992)}
\lref\JBYO{J. de Boer and Y. Oz, ``Monopole Condensation and Confining
Phase of N=1 Gauge Theories Via M Theory Fivebrane'', hep-th/9708044,
Nucl.Phys. {\bf B511} (1998) 155-196}
\lref\jlou{J. Louis and A. Micu,``Type II Theories Compactified on
Calabi-Yau Threefolds in the Presence of Background Fluxes'', hep-th/0202168}

\newsec{Introduction}

It was conjectured in \civ\ that large $N$ dual of $U(N)$
$\N=2$ gauge theory deformed by certain superpotential terms
is realized as type IIB string on Calabi-Yau threefolds with fluxes.
The evidence for this conjecture was provided by checking that
the low energy dynamics on both sides agree, at least up to the
order checked.  Namely the Calabi-Yau geometry led to a superpotential
for the gaugino fields, whose extremization yielded information
about the low energy dynamics.  This was checked using the gauge
theory analysis beginning with the exact $\N=2$ answer and
studying its deformation under the addition of superpotential.
The two objects look rather different.  On the gauge
theory side one studies the Seiberg-Witten curve
and its factorization locus, and on the geometry side
one studies extremization of a superpotential.
 The agreement for the
low energy dynamics (for example the tensions of the domain walls)
was checked to some order in a series expansion.  It is
natural to ask how to prove this equivalence to all orders,
which may also shed light on what it means to consider
a factorization locus of the Seiberg-Witten curve from the $\N=1$
perspective.

In this paper we find a proof of this equivalence.  The idea
is to relate the extremization of the superpotential to the
existence of some meromorphic function on the Riemann surface
with suitable divisors.  This in turn is equivalent
to specializing to the appropriate factorization locus
of the Seiberg-Witten curve.

We also push this idea further and recover the full
$\N=2$ low energy dynamics for $U(N)$ gauge theory by considering
a superpotential of degree $N+1$ and considering the locus
where $U(N)$ is broken down to $U(1)^N$.  By turning off
the superpotential we go back to a point on the Coulomb branch
of the $\N=2$ theory, and we are able to obtain the full
low energy dynamics of the $\N=2$ theory from the Calabi-Yau
geometry with fluxes.  It is quite interesting that
in the limit of turning off the superpotential
Calabi-Yau threefold becomes  the product of an $A_1$ geometry
with the complex plane, as is expected based on the
enhanced supersymmetry.
Nevertheless the information of the $\N=2$ low energy dynamics
survives in this limit.  For example the gauge coupling constants
are given by ratios of the periods of the Calabi-Yau threefold.
Even though the periods vanish in this limit, the ratios are finite
and yield the $\N=2$ low energy gauge couplings.

The organization of this paper is as follows:  In section 2 we
discuss the gauge theory analysis.  In section 3 we recall the
geometric dual and present a proof of its equivalence
with the gauge theory prediction.  In section 4 we show how
to recover the full $\N=2$ geometry from this setup.  Some technical
aspects of the computation are discussed in the appendices A,B and C.

\newsec{Field theory analysis}

In this section we will review the analysis of \civ\ giving rise to the exact
low energy superpotential of pure $\N=2$ $U(N)$
Yang-Mills theory deformed to $\N=1$ by a tree level
superpotential for $\Phi$ given by,
\eqn\defor{ W_{\rm tree} = \sum_{i=1}^{n+1} g_i u_i}
where $u_i={1\over i}$Tr$\Phi^i$.

The solution of this model is achieved by using the Seiberg-Witten curve of
the original $\N=2$ theory and going to the points on the Coulomb branch
where the susy vacua are not lifted by \defor.  As we
will review below, this approach reduces the problem
of finding the low energy superpotential $W_{\rm low}$, that is only a function
of $g_i$'s and $\Lambda$ (the scale of the $\N=2$ theory), to a well posed
factorization problem of a polynomial of degree $N$.
Note that one disadvantage for this method is that there is no
direct way to integrate in the gaugino superfields
which are important in the low energy dynamics
of the IR $\N=1$ theory.  This disadvantage
is resolved in the geometric dual description that we will review
in the next section.

% We will set $g_{n+1}=1$. Notice that for
% $n\neq 2$ is it easy to introduce $g_{n+1}$ back by dimensional analysis.

Classically, the vacuum structure of the theory is very simple. Solutions
to the $F$ and $D$-terms equations are given by $\Phi$ being diagonal with
eigenvalues solutions of,
$$ W'(x) = g_{n+1}x^n+\ldots + g_1 = g_{n+1}\prod_{i=1}^{n}(x-a_i) = 0.$$
The different vacua are given by the different choices of the number $N_i$
of eigenvalues of $\Phi$ equal to $a_i$. This is subject to the condition
$\sum_{i=1}^nN_i = N$ and the gauge group $U(N)$ is broken down to
$U(N_1)\times \ldots \times U(N_n)$.  Thus in the IR we end up
with pure $\N=1$ Yang-Mills theory with the group
$U(N_1)\times \ldots \times U(N_n)$.

In the Coulomb branch, the $\N=2$ theory is described at low energies by
an $U(1)^N$ effective theory. All the relevant quantum
corrections in the IR are given in terms
of an auxiliary Riemann surface and the periods of a particular meromorphic
one form.

The SW curve for a pure $U(N)$ gauge theory is given by \swc,
$$ y^2 = P_N(x)^2 -4\Lambda^{2N}$$
where $P_N(x, u_k) = <{\rm det}(xI-\Phi )>$ and
$u_k ={1\over k}{\rm Tr}\Phi^k$.

Once the tree level superpotential is introduced, all points in the Coulomb
moduli space are lifted except those for which $N-n$ mutually
local magnetic monopoles become massless. The
presence of the superpotential produces a condensate of monopoles and the
Higgs mechanism in the magnetic theory gives the expected
 confinement of the electric
$\N=1$ theory. Those points are where $<u_k>$'s are solution to,
\eqn\con{ P_N(x)^2 -4\Lambda^{2N} = F_{2n}(x)H^2_{N-n}(x)}
where $F_{2n}(x)$ and $H_{N-n}(x)$ are at this point arbitrary polynomials
with simple zeroes of degrees $2n$ and $N-n$ respectively. The fact that
$H^2_{N-n}(x)$ appears in the above signifies the appearance of $N-n$ mutually
local massless magnetic monopoles.
 From the
original $U(1)^N$ only $U(1)^n$ remains unbroken and the corresponding
coupling constants are given by the period matrix of the reduced curve,
$$ y^2 =  F_{2n}(x).$$
These $U(1)^n$ can also be thought of as $U(1)\subset U(N_i)$ for
$i=1,\ldots ,n$ from the classically
 unbroken group. On the other hand the pure $\N =1$ $SU(N_i)$ piece confines
in the IR, has a mass gap and gaugino condensation, i.e, $<{\rm
Tr}_{SU(N_i)}W^{\alpha}W_{\alpha}>\neq 0$.

For $U(N)$ the Coulomb moduli space has dimension $N$, parametrized for
example by the roots of $P_N(x)$. The condition \con\ implies that
$N-n$ of those have to be tuned in order to produce the $N-n$ double
roots on the RHS.
This implies that \con\ is satisfied on a codimension $N-n$ subspace of
the Coulomb moduli space.  Thus the factorization
condition \con\ does not lead
to a unique answer and there is an $n$ parameter family of
such factorizations.

Thus, for this subspace $<u_k>$'s
 are functions of $n$ parameters. Plugging this in the
superpotential $W_{\rm tree}$ an effective superpotential is obtained for
those $n$ variables,
\eqn\var{ W_{\rm eff} = \sum_{k=1}^{n+1}g_k <u_k>.}
Using the field equations from varying \var\ with respect to the $n$
variables one could get all $<u_k>$'s as functions only of $g_i$'s and
$\Lambda$. Substituting back in $W_{\rm eff}$ one gets $W_{\rm low} =
W_{\rm low}(g_i,\Lambda )$.

However, it is possible to restate this latter extremization
problem also in purely algebraic terms. In
\civ\ it was shown that extremizing the effective superpotential is
equivalent to imposing, (for a review of the proof see appendix A)
$$g_{n+1}^2 F_{2n}(x) = W'(x)^2 + f_{n-1}(x)$$
where $f_{n-1}(x)$ is a polynomial of degree $n-1$ completely fixed by
\con\ as we will show.

Putting these two factorizations together
 we thus have a purely algebraic description
of the low energy dynamics of the $\N=1$ theory.  The claim
is that we end up with the following
problem which is well posed and has a unique answer:
Find $P_N(x)$ such that,
\eqn\problem{P_N^2(x) - 4\Lambda^{2N} = {1\over g_{n+1}^2}
(W'(x)^2+f_{n-1}(x))H^2_{N-n}(x)}
where $W'(x)=g_{n+1}\prod_{i=1}^n (x-a_i)$ is given,
together with the following
condition,
$$ P_N(x)\to \prod_{i=1}^n (x-a_i)^{N_i} \quad {\rm as} \quad \Lambda \to 0$$

It is interesting to notice that these polynomials are a generalization of
Chebyshev polynomials that are the solution to the problem for $n=1$. The
proof that the solution to \problem\ is unique is given in appendix B.

Once $W_{\rm low}(g_r,\Lambda ) =
\sum_{r=1}^{n+1}g_i<u_i>$ is obtained, the following
information can be computed,
\eqn\vevs{ {\del W_{\rm low}\over \del g_r} = <u_i> \quad {\rm and} \quad
{\del W_{\rm low}\over \del {\rm log}\Lambda^{2N}} = <S_1+\ldots +S_n> }
where $S_i \equiv {\rm Tr_{SU(N_i)}W_{\alpha}W^{\alpha}}$ are the glueball
superfields of each $SU(N_i)$ factor.

It is possible to show that $-4 g_{n+1}<S_1+\ldots +S_n>$
is equal to the
coefficient of the $x^{n-1}$ monomial of $f_{n-1}(x)$ in \problem.
This fact plays an important role in section 3 and its proof is given at
the end of appendix A.

\newsec{Geometric dual analysis}

In \civ\ a geometric dual to the field theory in the previous section was
given. The dual theory was conjectured to have all the IR holomorphic
information of the original theory. More explicitly, the coupling constants
of the $U(1)$ factors and the effective superpotential for gaugino fields.
These conjectures were tested in a semi-classical series expansion up to
several orders.

In this section we will provide the proof that the gauge
theoretic prediction for the low
energy (holomorphic) dynamics is in exact agreement with the
geometric prediction. The
dual geometric description has the advantage
of also providing the effective superpotential for gaugino fields.

First a review of the geometric construction is given in order to set the
notation and then we show how the effective superpotential
$W_{\rm eff}(S_k)$ proposed in \civ\ gives equations whose solution is
completely equivalent to solving the problem proposed in the previous
section \problem.

\subsec{Review}

The starting point is to geometrically engineer the $\N=2$ $U(N)$ field
theory deformed by the superpotential term
\defor\ as the theory living on the world volume of D5
branes wrapping two cycles. We consider IIB
string theory on a non-compact Calabi-Yau 3-fold. The 3-fold is
a fibration of an $A_1$ ALE space over a complex
plane with $D5$ branes wrapping the nontrivial $S^2$ in the blown up $A_1$
singularity. At $n$ isolated points the Calabi-Yau 3-fold thus constructed is
singular and can be smoothed out by blowing up $S^2$'s or $S^3$'s.
Let us discuss this geometry in more detail.

The geometry corresponding to the theory without superpotential, i.e., to
the $\N=2$ theory is a product space of a complex plane with coordinate $x$
and the $A_1$ ALE space,
$$ uv + w^2 = 0.$$
In \ckv\ it was shown that adding the tree level superpotential \defor\ to the
field theory can be accounted for by allowing a nontrivial fibration given by,
\eqn\fibra{ uv + w^2 + W'(x)^2 = 0}
where $W'(x)=g_{n+1}\Pi_{i=1}^{n} (x-a_i)$. At each point $x=a_i$ there
is a blown
up $S^2$ and $N_i$ D5-branes wrapping around the $S^2$.

The dual theory proposed in \civ\  is obtained
via a geometric transition (as a generalization
of the $n=1$ case in \vafa ).
 The transitions takes place
when the $S^2$'s are blown down and $S^3$'s are blown up. The $N_i$ D5 branes
wrapping the $S_i^2$ located at $x=a_i$ disappear and get replaced by $N_i$
units of $H_{RR}$ flux through the new non-trivial $S^3_i$.

The transition to $S^3$'s corresponds to a complex deformation of the
geometry. The allowed deformations are computed by taking into account a
normalizability condition. The volume of a minimal lagrangian 3-cycle is
given by the absolute value of the integral of the holomorphic 3-form over
the cycle. In the non-compact geometry there are non-compact 3-cycles
$B_i$ whose volumes are infinite and need a large distance cut off
$\co$. The deformations that will correspond to dynamical fields are those
for which the corresponding variation of the holomorphic
form integrated over cycles will not depend on the cutoff
$\co$.    This is needed for the mode to be localized.
In other words,
\eqn\norcon{ {\rm lim}_{\co \to \infty}{\del \over \del b_k}\int_{B_i} \Omega }
is finite, where $b_k$'s are the coefficients of the deformation.
Actually we also allow log normalizable, i.e. allow
divergence of the form ${\rm log}\co$.
This is deeply connected with asymptotic freedom of the
underlying gauge theory.
This condition fixes the form of the possible complex
 deformations of
\fibra\ to be,
$$ uv + w^2 + W'(x)^2 + f_{n-1}(x) = 0$$
where,
$$f_{n-1}(x) =  \sum_{j=0}^{n-1}b_j x^j.$$
The variation of $b_{n-1}$ term corresponds to log
normalizable term.
Type IIB on this geometry
gives rise to an effective $\N=2$ $U(1)^n$ field theory
in four dimensions. However, the presence of fluxes induces electric and
magnetic FI terms in the effective action allowing for a spontaneous
symmetry breaking to $\N=1$.

The effective superpotential for Calabi-Yau 3-folds with
fluxes was considered in \tva\mayr\
(see also the more recent work \jlou\ ).
This is given by
$$ W_{\rm eff} = \int_{CY} H\wedge \Omega  $$
where $H=H_{RR}-\tau_{\rm IIB}H_{NS}$ and $\Omega$ is the holomorphic three
form of the CY 3-fold.

Let us choose a symplectic basis for three cycles $A_i$ and $B_i$, with $A_i$
identified with the blown up $S^3_i$ and $B_i$ with the dual non-compact
cycle to $S^3_i$. In terms of this basis the superpotential
corresponding to the classical vacuum\foot{We assume that $N_i$'s do not
have a common factor} where
$N=\sum_{i=1}^n N_i$, is given by,
\eqn\simpl{ W_{{\rm eff}} =\sum_{i=1}^n\left( \int_{A_i} H\int_{B_i}
\Omega -
\int_{B_i}H\int_{A_i}\Omega \right).}

Using the fact that the D5-branes have been replaced by fluxes we get,
\eqn\proH{ \int_{A_i} H = N_i  \quad  {\rm and} \quad
\int_{B_i} H  =\tau_{\rm YM}\quad {\rm for} \quad i,j = 1,\ldots ,n. }
The second condition implies that $\int_{B_i}H$ is a constant
independent of $i$, and thus $\int_{B_i-B_j}H=0$.  Note that
 since $B_i$ cycles are non-compact $\int_{B_i} H$ 
 is actually infinite. This IR divergence can be traced back to the
original Yang-Mills UV divergence. This is dealt with by the introduction of
a cut off $\co$. Following the same steps we can identify the constant with
$\tau_{\rm YM (\co )}$, the bare Yang-Mills coupling
as was done in \vafa.

Plugging this in the superpotential \simpl,
\eqn\dualsup{ W_{\rm eff} =\sum_{i=1}^n N_i \Pi_{i} +\tau_{\rm YM (\co
)} \sum_{i=1}^n S_i  }
where, $S_i \equiv \int_{A_i}\Omega$ and $\Pi_i \equiv \int_{B_i}\Omega$.

The $S_i$ and $\Pi_i$ period integrals can be shown to reduce to line
integrals over the complex $x$ plane of the following effective one form,
\eqn\effone{\lambda_{\rm eff} = \sqrt{W'(x)^2+ f_{n-1}(x)}dx.}

There are $2n$ branch points on the x-plane with $n$ branch cuts running
between pairs as shown in Figure 1. $S_i$'s are integrals of
$\lambda_{\rm eff}$ around the i-th branch cut, $\alpha_i$. On the other
hand, $\Pi_i$'s are integrals from $x=\co$ on the lower sheet to $x=\co$ on
the upper sheet following $C_i$'s.

Adding the contours of all $S_i$'s and deforming it to enclose $x=\infty$,
it is easy to show that,
$$\sum_{i=1}^n S_i= -{1\over 4g_{n+1}} b_{n-1}$$
by computing the residue of the pole at infinity.

Therefore, the superpotential can be written as,
$$W_{\rm eff} = \sum_{i=1}^n N_i \Pi_i - \tau_{\rm YM (\co
)}
{1\over 4g_{n+1}}b_{n-1}. $$

The effective superpotential is only a function of $S_i$'s for $\Pi_i =
{\del \F \over \del S_i}$, where $\F=\F (S_1,\ldots ,S_n)$ is the prepotential
of the CY 3-fold. The field equations are given by,
$$ {\del W_{\rm eff}(S_k)\over \del S_i} = 0 \quad {\rm for}
\quad i=1,\ldots ,n. $$
However, it turns out to be more useful to use a change of variables from
$\{ S_1,\ldots , S_n\}$ to $\{ b_{n-1}, \ldots ,b_0\}$, which is generically
non-singular.

With the change of variables, the field equations are given by,
\eqn\fieq{ \sum_{i=1}^n N_i{\del \Pi_i \over \del
b_{n-1}} -  {\tau_{\rm YM (\co
)} \over 4g_{n+1}}
= 0  \quad {\rm and} \quad \sum_{i=1}^n N_i{\del \Pi_i \over \del
b_{j}}=0 \quad {\rm for} \quad j=0,\ldots ,n-2 .}

%%%%%%%%%%%%%%%%%%%%%%%%%%%%%%%%%%%%%%%%%%%%%%%%%%%%%%%%%%%%%%%%
\bigskip
\centerline{\epsfxsize=0.65\hsize\epsfbox{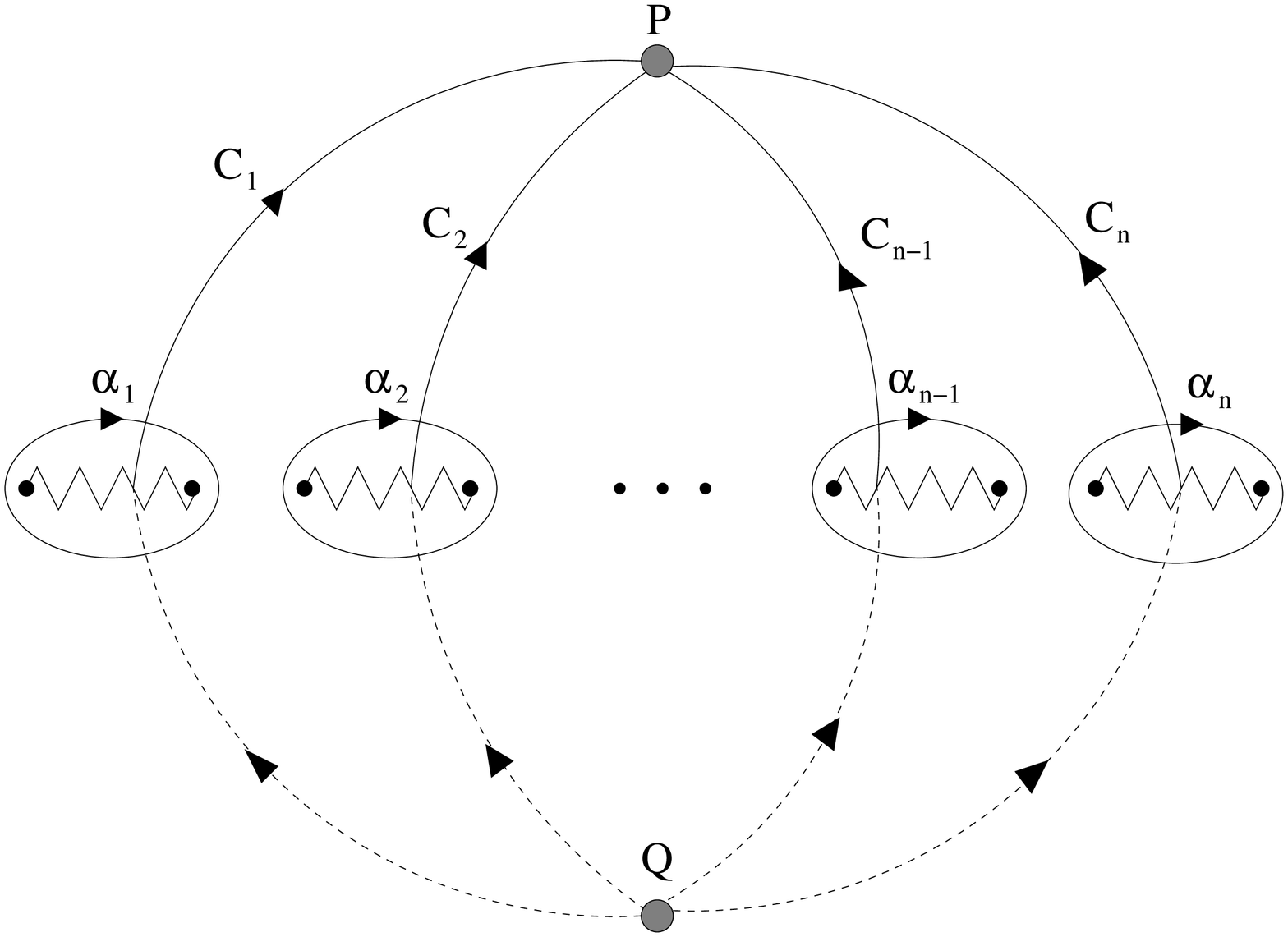}}
\noindent{\ninepoint\sl \baselineskip=8pt {\bf Figure 1}:{\sl Contours of
 integration. The points $P$ and $Q$ represent $\co$ on the upper and lower
sheets of the Riemann surface.}}
\bigskip

%%%%%%%%%%%%%%%%%%%%%%%%%%%%%%%%%%%%%%%%%%%%%%%%%%%%%%%%%%%%%%%%

\subsec{Conjectures}

Let us recall the conjectures made in \civ. Consider the original $\N=2$
$U(N)$ theory in the classical vacuum where $U(N)$ is broken down to
$U(N_1)\times \ldots \times U(N_n)$. As mentioned in section 2, each factor
$U(N_i)=U(1)\times
SU(N_i)$ in the IR is expected to give a free $U(1)$ and gaugino condensate
for the confining $SU(N_i)$ piece, i.e.,
$<{\rm Tr}_{SU(N_i)}W^{\alpha}W_{\alpha}>\neq 0$. The holomorphic
information, as mentioned before, is composed of the coupling constants $\tau_{ij}$ of
the $U(1)^n$ factors and the effective superpotential for the glueball
fields $S^{\rm gf}_i ={\rm Tr}_{SU(N_i)}W^{\alpha}W_{\alpha}$ where $({\rm
gf})$ stands for glueball field.

The duality map is the following: the $\N=2$ $U(1)^n$ vector superfields in
the Calabi-Yau with blown up $S^3$ that can be decomposed in $\N=1$
superfield notation as $(W^i_{\alpha}, S_i)$, are identified with the
$U(1)^n$ $W^i_{\alpha}$ and $S^{\rm gf}_i$ of the original theory
respectively.
Namely, the lowest component of the glueball field $S_i^{\rm gf}$ is the
holomorphic volume of the $S^3_i$, i.e, $S_i=\int_{A_i}\Omega$.

With this identification two physical predictions
follow which can be stated as mathematical
conjectures, namely

{\bf Conjecture 1}: The coupling constants $\tau_{ij}$ of the original
$U(1)^n$ groups are given in the dual geometry by,
$$ \tau_{ij} =\left. { \del^2 {\cal F}\over \del S_i \del S_j}\right|_{S_k
\to <S_k>} $$
where $<S_k>$ are the expectation values of the massive $S_i$ fields.
More precisely, in the original field theory, the overall $U(1)\subset
 U(N)$ decouples from the other $U(1)^{n-1}$'s. In an
appropriate basis the
 couplings are given by,
$$ \tau_{ij} \quad {\rm with} \quad i,j=1,\ldots ,n-1 \qquad \tau_{i,n}=0
\quad {\rm for} \quad i=1,\ldots ,n
\qquad {\rm and} \qquad \tau_{nn} =\tym $$
where $\tau_{ij}$ is the period matrix of the reduced SW curve,
$y^2= W'(x)^2 + f_{n-1}(x)$ solution to \problem.

{\bf Conjecture 2}: Solving the problem \problem\ to find $f_{n-1}(x)$ and
$<{\rm Tr} \Phi^k>$ for $k=1, \dots , n+1$ is equivalent to solving the field
equations \fieq\ arising from the dual effective superpotential \dualsup.
In particular,
$f_{n-1}(x)$ appearing in the geometry is the same as that appearing in the
field theory and
$$W_{\rm eff}(<S_i>) = W_{\rm low}(g_i, \Lambda).$$

In the next section we will give first the proof to conjecture 2 and then
using the relation between the geometries conjecture 1 will be shown to follow.

\subsec{Proof of conjectures}

Consider the effective superpotential,
$$ W_{\rm eff} = \int_{\rm CY} H\wedge \Omega $$
where the CY 3-fold is given by, $ uv + w^2 + W'(x)^2 + f_{n-1}(x) = 0$.
Recall that $f_{n-1}(x) = b_{n-1}x^{n-1}+\ldots + b_0$.

The field equations are obtained by varying with respect to the
deformations $b_k$,
$$ {\del W_{\rm eff} \over \del b_k} = \int_{\rm CY} H\wedge
{\del \Omega \over \del b_k } = 0.$$

After integrating over the quadratic pieces in the geometry the integral
over the CY-3-fold is reduced to an integral over a Riemann surface
$\Gamma$,
\eqn\soH{ y^2 =  W'(x)^2 + f_{n-1}(x) }
There are two special points on $\Gamma$ for our discussion, they are
located at the two pre-images of $\infty$ of
$x$. Let us denote them by $P$ and $Q$.

In section 3.1 we denoted the reduction of $\Omega$ to a one form over
$\Gamma$ by,
$$ \lambda_{\rm eff} = \sqrt{W'(x)^2+f_{n-1}(x)}dx. $$
Let us also introduce a one form $h$ for the reduction of $H$,
$$ h = \int_{S^2} H .$$
Note that $h$ is subject to constraints coming from \proH, namely,
\eqn\condh{ \oint_{\alpha_i}h = N_i \quad {\rm and} \quad 
\int_{C_i}h =\tau_{\rm YM} \rightarrow \oint_{C_i-C_j}h = 0 }
for all $i$ and $j$ in $\{1,\ldots ,n\}$. See figure 1 for the definition
of $\alpha_i$'s and $C_i$'s.

Moreover, it is clear by adding up the $\alpha_i$ contours that,
\eqn\PQ{ \oint_{P}h = N   \quad {\rm and }  \quad \oint_{Q}h = -N ,}
where the integrals run over a path enclosing $P$ and $Q$ respectively.
Therefore $h$ should have precisely a pole of order 1 at $P$ and $Q$ with
residue $N$ and $-N$ respectively.

On the Riemann surface the extremization
of the superpotential gives
\eqn\hol{ \int\!\!\int_{\Gamma}h\wedge {\del 
\lambda_{\rm eff}\over \del b_k} = 0 \quad
{\rm for} \quad k=0,\ldots, n-2, n-1.}
Notice that ${\del \lambda_{\rm eff}\over \del b_k}$ for $k=0,\ldots ,n-2$
are holomorphic one forms on $\Gamma$. These form a complete
 basis
of holomorphic one forms. Therefore by the Riemann
bilinear identities \hol\ is satisfied if and only if
 $h$ is a holomorphic one form on
$\Gamma -\{ P,Q\}$.  This will also make the equation
for varying of $b_{n-1}$ satisfied on $\Gamma -\{P, Q\}$.
But for all $b_k$ variations
we also need to consider the potential contribution of the
integral \hol\ from $P,Q$.  By using the Riemann bilinear
identity this is equivalent to the contribution
\eqn\poles{\oint_P h\int_P^Q {\del 
\lambda_{\rm eff}\over \del b_k}-\oint_P {\del 
\lambda_{\rm eff}\over \del b_k}\int_P^Q h =0.}
Using \PQ\ and \condh\  we can write this as
\eqn\fpoles{N\int_P^Q {\del 
\lambda_{\rm eff}\over \del b_k}-\oint_P {\del 
\lambda_{\rm eff}\over \del b_k}\tau_{\rm YM} =0.}
For $k=1,...,n-2$ the second term vanishes
because $\omega_k={\del \lambda_{\rm eff}\over \del b_k}$ is a
holomorphic one form. Thus we obtain
\eqn\holpart{N\int_P^Q \omega_k=0.}
Note that this is well defined up to addition 
of periods, depending on which path
one takes from $P$ to $Q$.
This equation implies, according to Abel's theorem
that there must be a meromorphic function on $\Gamma$
with an $N$-th order zero on $P$ and an $N$-th order pole on $Q$.
For $k=n-1$, since $\omega_{n-1}={1\over g_{n+1}}{\del 
\lambda_{\rm eff}\over \del b_{n-1}} \sim dx/x$ as $x\rightarrow \infty$,
 we have to introduce
a cutoff $\Lambda_0$, as discussed before.  We obtain
$$N\int_P^Q \omega_{n-1}-\tau_{\rm YM}=0$$
where the first term gives $2N  {\rm log}[\Lambda/\Lambda_0] $
for some $\Lambda$ (depending on $b_i$) and we obtain
\eqn\couy{\int _P^Qh= \tau_{\rm YM}=2N{\rm log}[\Lambda/\Lambda_0]}

Now that we have translated the field equations \hol\ into the existence of
a holomorphic one form $h$ on $\Gamma$ with certain properties, and
the existence of a meromorphic function with divisor $N[P-Q]$
 the final
step is to find $f_{n-1}(x)$ such that $\Gamma$ defined by \soH\ admits such
an $h$ and such a meromorphic function.

We will now show that these exist if $f_{n-1}(x)$ is such that the
following is true,
\eqn\cond{ \left( W'^2(x)+f_{n-1}(x) \right) H^2_{N-n}(x)
= g^2_{n+1}(P_N(x)^2 -\gamma^2)  .}
for some $H_{N-n}(x)$ and $P_N(x)$, where
$P_N(x)\to \prod_{i=1}^n (x-a_i)^{N_i}$ as $\gamma\to 0$.
The factor $g_{n+1}$ is introduced only to normalize the coefficient
of $x^N$ in $P_N(x)$ to one.

Consider, the function $z$ on $\Gamma$,  defined by,
$$ z = P_N(x) - {1\over g_{n+1}} y H_{N-n}(x).$$
Note that due to \cond, $z$ satisfies the following equation
on $\Gamma$:
\eqn\anver{z-2P_N(x)+{\gamma^2\over z}=0. }
$z$ has a zero of order $N$ at $P$ and a pole of order $N$ at
Q. This is one condition we needed to satisfy.
Moreover, $z$ does not have any zeros or poles in $\Gamma -\{P,Q\}$.
This follows from \anver .
This implies that ${1\over 2\pi i}{dz\over z}$ satisfies \PQ. We claim that,
$$ h ={1\over 2\pi i}{dz\over z}.$$
We need to check that the conditions \condh\ are satisfied.
In order to compute the periods
of $h = {1\over 2\pi i}{dz\over z}$
over $\alpha_k$'s
notice that the answer is independent of $\gamma$.  This is
because $z$ is a well defined function on $\Gamma -\{ P, Q\}$,
and its phases can change only by an integer multiple of $2\pi i$.
This implies that the
evaluation can be performed in the limit $\gamma \to 0$.
$$ \oint_{\alpha_k} {dz\over z} = \oint_{\alpha_k} d(\log z) =
\oint_{\alpha_k}d(\log ( 2P_N(x)|_{\gamma\to 0})).$$
But $P_N(x)|_{\gamma\to 0} = \prod_{j=1}^n (x-a_j)^{N_j}$ and therefore,
$$ \oint_{\alpha_k} h ={1\over 2\pi i}\oint_{\alpha_k} {dz\over z} =
N_k {1\over 2\pi i}\oint_{\alpha_k}d(\log (x-a_i)) = N_k.$$
We can also compute $\int_{C_i-C_j}{dz\over z}$. This can be
done using the same argument as before and realizing that
going around the $C_i-C_j$
cycle we do not cross any branch cut of the $\log(x-a_i)$
functions. Hence,
$$ \oint_{C_i - C_j} {dz\over z} = 0.$$
%

%%%%%%%%%%%%%%%%%%%%%%%%%%%%%%%%%%%%%%%%%%%%%%%%%%%%%%%

Finally, we have to check \couy\ which
relates $\gamma$ to the parameters of the original
Yang-Mills theory. 
{}From the definition of $z$ we see that the $\gamma$ gets
identified with $\gamma = \pm 2\Lambda^{N}$.

In order to complete the proof of the second conjecture we only have to
show that
$$W_{\rm eff}(<S_k>)=W_{\rm low}(b_k, \Lambda).$$

{}The final result in section 2, showed from field theory that,
$$ {\del W_{\rm low}\over \del {\rm log}\Lambda^{2N}} = -{1\over 4g_{n+1}}
b_{n-1}.$$
where $b_{n-1}$ is the coefficient of $x^{n-1}$ in $f_{n-1}(x)$ from field
theory. However, we also have that,
$${\del W_{\rm eff}(<S_k>)\over \del {\rm log}\Lambda^{2N}} =
-{1\over 4g_{n+1}}b_{n-1}$$
where $b_{n-1}$ is the coefficient of $x^{n-1}$ in $f_{n-1}(x)$ from the CY
3-fold. Given that we have shown that the two polynomials
$f_{n-1}(x)$ are equal, the
following is true,
\eqn\gian{{\del W_{\rm eff}(<S_k>)\over \del {\rm log}\Lambda^{2N}} = {\del W_{\rm
low}(b_k's, \Lambda )\over \del {\rm log}\Lambda^{2N}}.}

Finally, showing that $\left. W_{\rm eff}(<S_k>)\right|_{\Lambda\to 0}$ is
equal to $W_{\rm low}(g_k, \Lambda\to 0)$ will complete the proof.
{}From section 2, taking $\Lambda \to 0$ is the classical limit and
$$W_{\rm low}(g_k, \Lambda\to 0) = \sum_{i=1}^n N_i \sum_{k=1}^{n+1}
{1\over k}g_k a_i^k .$$
On the other hand, from the geometry, setting $\Lambda$ to zero gives
$f_{n-1}(x)=0$ and
the effective one form simplifies $\lambda_{\rm eff} = ydx = W'(x)dx$.
Taking into account that
$b_{n-1}$ goes to zero as a polynomial in $\Lambda$ we get that the second
term in the effective superpotential \dualsup\ given by $\tym <S_1
+\ldots +S_n>$ goes to zero in the limit. Notice that we could add to
our definition of $W_{\rm eff}$ \dualsup\ an arbitrary function of the form,
$$ (N_1+N_2+\ldots + N_n)G(g_k's, \co ) $$
which is $\Lambda$ independent and does not affect the validity
of \gian .
Such an addition does not have any effect on physical quantities. 
It does not affect the field equations because it is an additive
constant to the superpotential.
Having shown that such an addition is
harmless, let us choose $G(g_k's, \co ) = W(\co )$.
Therefore taking the limit $\Lambda \to 0$ we get,
$$ W_{\rm eff}(<S_k>) = -\sum_{i=1}^{n}N_i \int_{a_i}^{\co} W'(x)dx + N
W(\co )= \sum_{i=1}^n N_i \sum_{k=1}^{n+1}
{1\over k}g_k a_i^k.$$
This completes the proof of the second conjecture.

\bigskip
{\bf Coupling constants for $U(1)^n$}

In order to establish conjecture 1 we only have to show that the couplings
in the dual theory given by,
$$ \tau_{ij} = {\del^2 {\cal F}\over \del S_i \del S_j}$$
in some appropriate basis decompose into the period matrix of the
auxiliary Riemann surface $\Gamma$ and the coupling of original $U(N)$
theory $\tym$.

The change of basis is easy to guess if we look at the field equations from
\dualsup,
$${\del \over \del S_j}\sum_{i=1}^n N_i \Pi_i + \tym = 0  \quad {\rm for}
\quad j=1,\ldots ,n$$
using that $\Pi_i = {\del {\cal F}\over \del S_i}$ the equations can be
written as,
$${\del \over \del S_j}\sum_{i=1}^n N_i{\del \over \del S_i} {\cal F} =
-\tym. $$

{}From this it is natural to define basis $\{ S_{12},S_{23}, \ldots ,
S_{n-1,n}\; ,S_+\}$ such that,
$$ {\del\over\del S_+} = \sum_{i=1}^n N_i {\del\over\del S_i} \qquad {\rm
and} \qquad
{\del\over\del S_{i,i+1}} = {\del\over\del S_i} -{\del\over\del S_{i+1}}$$
In this new basis the field equations become,
$$ \tau_{i,+} = {\del^2 \over\del S_{i,i+1}\del S_+}{\cal F} = 0
\quad {\rm for} \quad i = 1, \ldots ,n-1 \qquad {\rm and } \qquad
\tau_{++} = {\del^2 \over\del S_+^2}{\cal F} = -{1\over N}\tym. $$

Finally, we only have to show that the remaining elements of $\tau_{ij}$
give the period matrix of $\Gamma$.
Consider,
\eqn\coupl{ \tau_{ij} = {\del^2 \over \del S_{i,i+1}\del S_{j,j+1}}{\cal F} =
{\del \over \del S_{i,i+1}}\left( \Pi_j - \Pi_{j+1} \right) }
{}From figure 1 it is clear that $\Pi_j-\Pi_{j+1} = \int_{C_j -
C_{j+1}}y dx$ is an integral over a compact cycle.
One more change of variables is needed. Let the new independent variables be
$\{ b_0, \ldots, b_{n-1}\}$. Using this, \coupl\ becomes,
\eqn\matt{\tau_{ij} = \sum_{k=0}^{n-2}{\del b_k \over \del S_{i,i+1}}
{\del \over \del b_k}\left( \int_{C_j -
C_{j+1}}y dx \right) +{\del b_{n-1} \over \del S_{i,i+1}}
{\del \over \del b_{n-1}}\left( \int_{C_j -
C_{j+1}}y dx \right).}
However, recalling that $b_{n-1} = -4(S_1+\ldots + S_n) = -4S_+$
the second term drops out since ${\del b_{n-1} \over \del S_{i,i+1}}=0$.

{}Using that $y^2 = W'(x)^2 + b_{n-1}x^{n-1}+\ldots +b_1 x+ b_0$ it is easy
to see that,
$${\del \over \del b_k} (y dx) = {1\over 2}{x^{k}\over y}dx \quad {\rm for}
\quad k=0,\ldots , n-2$$
forms a basis of holomorphic one forms over
$\Gamma$. Moreover, $S_{i,i+1} =\int_{\gamma_i}ydx$, with $\gamma_i$
integral linearly independent combinations of $\alpha_j$'s. Together,
$(\gamma_i , C_i-C_{i+1})$ form a basis for $H_1(\Gamma
,{\bf Z})$. Therefore, $\tau_{ij}$ given in \matt\ is a period matrix of
$\Gamma$.
This completes the proof
of the conjectures.

%%%%%%%%%%%%%%   Part II   %%%%%%%%%%%%%%%%%%%%%%%%%%%%%%%%%%%%%%%%%

\newsec{Derivation of the Seiberg-Witten solution for $\N=2$ $U(N)$ from
Fluxes}

The $\N=1$ theories we have studied up to now are deformations of
pure $\N=2$ $U(N)$ Yang-Mills. It is natural to ask to what extent one can recover information of
the original $\N=2$ theory as the deformation is turned off. This is
the main issue we want to address in this part of the paper.

The idea is to look for deformations $W_{\rm tree}$ that will provide
information about an arbitrary
Coulomb point of the original $U(N)$ theory. This will
turn out to be a potential of degree $N+1$,
$$ W_{\rm tree} = \sum_{k=1}^{N+1}
{g_k\over k}{\rm Tr}\Phi^k $$
and we consider the vacuum which breaks $U(N)$ to $U(1)^N$ generically. In
this vacuum $\Phi$ is given by ${\rm diag}(a_1, \dots ,a_N)$, where
$a_i$'s are defined by,
$$W'(x) = g_{N+1}x^N+\ldots +g_1 = g_{N+1}\prod_{k=1}^{N}(x-a_i).$$
The limit that allows us to get back to the $\N=2$ theory is $g_{N+1}\to
0$ while keeping $a_i$'s fixed. The $a_i$'s will correspond to a generic
point in the Coulomb branch of the $\N=2$ theory.
It is natural to suggest that all $\N=2$ information, if any, in the $\N=1$
theory will have to come from quantities that do not depend on $g_{N+1}$.
Moreover, intrinsically $\N=1$ objects like gaugino vev's will all vanish
as $g_{N+1}\to 0$.

\subsec{Seiberg-Witten Curve}

We will see first how the
$\N=2$ curve arises as a solution to the field equations \fieq\ of the
effective superpotential \dualsup\ for gaugino fields $S_i$'s.

Let us rewrite \dualsup\ using $N_i=1$ for $i=1,\ldots ,N$,
$$W_{\rm eff}(S_1,\ldots ,S_N) = \sum_{i=1}^N \Pi_i + \tym \sum_{i=1}^N S_i. $$
where $\Pi_i = {\del \F \over \del S_i}$ and $\F$ is the prepotential of
the CY 3-fold,
$$uv + w^2 + W'(x)^2+f_{N-1}(x)=0.$$
In this case the field equations arising from the superpotential are
hard to solve. The main problem being the determination of the prepotential
$\F$. However, they can be solved for any $N_i$'s in a semi-classical
expansion. See appendix C for examples.

Luckily, in this case the factorization problem \problem\ is trivial and in
section 3 we gave a general proof of the equivalence of the two. So we can
simply use \problem\ with $n=N$ to get,
$$ P^2_N(x) - 4\Lambda^{2N}  = {1\over g^2_{N+1}} \left( W'(x)^2+b_{N-1}(\Lambda )x^{N-1}+\ldots +b_0(\Lambda ) \right) .$$
{}From this we get that $b_{k}(\Lambda )=0$ for $k=1,\ldots ,N-1$ and
$b_0=- 4g_{N+1}\Lambda^{2N}$ is a solution. In appendix B we show that this is
indeed the unique solution. Therefore, $W'(x) = g_{N+1} P_N(x) =g_{N+1}
<{\rm det}(xI-\Phi)>$.

This implies that the vev's of the Casimirs $u_k = {1\over k}{\rm Tr}\Phi^k$
are not modified quantum mechanically and $<u_k> = (u_k)_{\rm class}$.
Let us check that this is indeed the case from the low energy
superpotential of the dual theory, i.e. $W_{\rm eff}(<S_1>,\ldots ,<S_N>)$.

The effective superpotential after minimization procedure can be used to
compute the quantum expectation value of the Casimir operator $<u_k>$ as
well as the expectation value of $<S>=<S_1+\ldots +S_N>$ as follows,
$$ {\del W_{\rm eff}\over \del g_k} = <u_k>  \quad {\rm and} \quad
{\del W_{\rm eff}\over \del {\rm Log}\Lambda^{2N}} = <S>$$
But we know from the solution to the field equations that the expectation
value of $<S>$ is zero because it is proportional to $b_{N-1}$. This
implies that $W_{\rm eff}(<S_1>,\ldots ,<S_N>)$ is not a function of
$\Lambda$ and
therefore it can be computed at any value, in particular, at $\Lambda
=0$. This implies that,
$$ W_{\rm eff}(<S_1>,\ldots ,<S_N>)  = W_{\rm class} (g_l) $$
Therefore,
$${\del W_{\rm eff}(g_l)\over \del g_k} = (u_k)_{\rm class} $$
as it should be consistent with the result from the curve.
Now recall that the geometry of the Calabi-Yau 3 fold after the transitions
is given by,
$$uv + w^2 + W'(x)^2+f_{N-1}(x) =0 .$$

Using the result of minimizing the superpotential we get,
$$ uv + w^2 + g_{N+1}^2\left( P_N(x)^2-4\Lambda^{2N}\right) =0.$$

There are several interesting observations to make from this:
Notice that the auxiliary Riemann surface $\Gamma$ used to compute
periods is exactly equal to the
Seiberg-Witten curve for pure $\N=2$ $U(N)$ after absorbing a factor of
$g_{N+1}$ in the definition of $y$.

This is surprising given that we expect to recover the $\N=2$ answer only
when $g_{N+1}$ is taken to zero. However, the SW curve is the solution to the
field equations for all $g_{N+1}$. Moreover, for $g_{N+1}\to 0$, the
geometry of the CY 3-fold reduces to that of an $A_1$ singularity
trivially fibered over the $x$-plane as expected from enhanced
supersymmetry in this limit.  This looks like the classical
limit of the $\N=2$ theory, and so one would like to
see how the exact quantum $\N=2$ answer is recovered.

Let us consider in more detail the way periods $S_i$'s and $\Pi_i$'s of the
holomorphic three form over $A_i$'s and $B_i$'s cycles depend on
$g_{N+1}$. As mentioned in section 3 the periods can be written as
integrals of an effective one form \effone\ over the x-complex plane.
$$ \lambda_{\rm eff} = \sqrt{W'(x)^2 +f_{N-1}(x)}dx = g_{N+1}\sqrt{P(x)^2
-4\Lambda^{2N}}dx $$
The contours of integration only depend on the position of the branching
points $a_i$'s. This implies that, ${1\over g_{N+1}}S_i$ and
${1\over g_{N+1}}\Pi_i$ are independent of $g_{N+1}$. 
The $\N=1$ fields $S_i$ and $\Pi_i$ go to
zero in the limit $g_{N+1}\to 0$.
Recall that the $U(1)^N$ couplings in the dual theory are given by,
$$ \tau_{ij} = {\del \over \del S_i}\Pi_j =
{\del \over \del \left({1\over g_{N+1}}S_i\right)}\left( {1\over g_{N+1}}\Pi_j\right)$$
and therefore are trivially $g_{N+1}$ independent.
Furthermore, as discussed in section 3, in a suitable basis
$\{ S_{12}, S_{23}, \ldots, S_{N-1,N}, S_+\}$ defined by,
$$ {\del \over \del S_{i,i+1}} = {\del \over \del S_i} -{\del \over \del
S_j} \quad {\rm and} \quad S_+ = S_1 + S_2 + \ldots +S_N$$
the $U(1)$ coupling
$\tau_{++} \equiv {\del^2 \F\over \del S_+^2}=-{1\over N}\tym$
decouples, i.e, $\tau_{+i}=0$ for $i=1,\dots ,N-1$ and
$\tau_{ij} \equiv {\del^2 \F\over \del S_{i,i+1}\del S_{j,j+1}}$ is equal
to the period matrix of $y^2 = P_N(x)^2-4\Lambda^{2N}$.

We have thus recovered the $U(1)^N$ coupling constants of the $\N=2$ theory
that are originally given by $\tau_{++} =-{1\over N}\tym$ and
$\tau_{ij} = {\del a_{Dj}\over \del a_i}$.

\subsec{$\N=2$ dyons}

The $\N=2$ data also contains information about the mass of BPS
particles. It is therefore interesting to see how this data comes out of our
$\N=1$ theory. At first sight this seems not to be possible given that in the
$\N=1$ theory  dyons are not BPS states.
 However we will see that the key is to realize that the dual
theory contains non zero fluxes of $H = H_{RR}+\tau_{\rm IIB} H_{NS}$
through 3-cycles. This three form carries nontrivial information
because, as we will show, its integral over $S^2$ in the fiber is $g_{N+1}$
independent. Of course, the
computation of the exact mass of the dyons is conceptually correct only
when $g_{N+1}\to 0$.

Let us start by identifying the electric and magnetic objects of
the $\N =2$ system before and after the transition.
Consider first the geometry before the transition with one $D5$ brane
wrapping each $S^2_i$, where $S^2_i$ is the non-trivial element in $H_2$ of
the $A_1$ fiber located at $x=a_i$. These are the points where the
holomorphic volume $\alpha_i$ is zero i.e. the solutions to the classical
field equations obtained from $W_{\rm tree}$. Here we are considering the case
where the kahler volume is also zero but the stringy volume is non zero due
to the contribution of $B_{NS}$.

If $g_{N+1}\to 0$ we expect the fibration to become
trivial. The geometry is just the product of the $x$-complex plane and the
$A_1$ ALE space where the singularity is resolved only by $B_{NS}$ and
$B_{RR}$. These fluxes are clearly constant. $D5$ branes are still
wrapping the $S^2$ at the same locations in the x-plane,
therefore we are at some point in
the Coulomb branch of the classical $\N =2$ $U(N)$ theory. At this generic
point the gauge group is broken down to $U(1)^N$. The electric and magnetic
particles can be easily identified as follows. $W$-bosons with charges
$(1,-1)$ under $U(1)_i\times U(1)_j$ are identified with open
strings stretching between $D5$ branes at $x=a_i$ and $x=a_j$. Given
that the fibration is trivial, the mass of such a string is simply its
tension times its length $m = |a_i-a_j|$. A magnetic object, on
the other hand, can be identified with a $D3$ brane wrapping a 3-chain
given by $S^2\times I_{ij}$, where $I_{ij}$ is the interval in the x-plane
from $a_i$ to $a_j$. Its mass is given by the volume of the 3-chain times
the tension of the brane. The mass is therefore given by
$m = \left| \int_{S^2}(\tau_{\rm IIB} B_{NS} + B_{RR})\right|
|a_i-a_j|$. Recall that the holomorphic coupling of the 4 dimensional
field theory is given by $\tau_{\rm YM} = \int_{S^2}(\tau_{\rm IIB} B_{NS}
+ B_{RR})$, therefore $m = |\tau_{\rm YM}(a_i-a_j)|$. This is also the
result from field theory classically.

It is important to keep in mind that $\tau_{\rm YM}$ is constant only in
the classical theory. Quantum mechanically we expect 
$B$ to vary over $x$-plane.  Thus the mass will be
given by an integral over the
path in the x-complex plane joining $a_i$ to $a_j$ times
the $B$-field over each point. Let us now write the
central charge $Z_m$ instead of the mass $m=\sqrt{2}|Z_m|$.
Therefore,
\eqn\cent{ Z_m = \int_{I_{ij}}\int_{S^2}B \wedge dx}
where we have defined $B= \tau_{\rm IIB} B_{NS}+ B_{RR}$.

As $g_{N+1}$ is turned on, the IR physics is described by the geometry after
the transition where D5 branes wrapping 2-cycles have been replaced by
fluxes over the new 3-cycles.

As the $S^2_i$ and $S^2_j$ are blown down the 3-chain $S^2\times I_{ij}$
on which the $D3$ was wrapped becomes a 3-cycle given by $B_i-B_j$. In
terms of the basic cycles in the auxiliary Riemann surface $\Gamma$ this is
the same as $\beta_i - \beta_j$ and the integral \cent\ can be written as,
$$Z_m  = \oint_{\beta_i}\int_{S^2}B \wedge dx - \oint_{\beta_j}\int_{S^2}B \wedge dx .$$
Integrating by parts in order to bring in $H=dB = H_{RR} + \tau H_{NS}$,
\eqn\ce{Z_m =  \int_{(\beta_i-\beta_j)\times S^2}x\ H.}
This formula for the BPS mass is only valid in the limit $g_{N+1}\to
0$. The reason being that only when the geometry becomes a trivial
fibration, the kahler volume of the $S^2$ and the complex volume of the
interval $I_{ij}$ combine.

We are only left with the computation of the one form $h\equiv \int_{S^2}H$
in the dual
geometry. But recall from section 3.3 that such a one form was found in the
general case to be given by \anver,
\eqn\swone{ h ={1\over 2\pi i}{dz\over z} \quad {\rm with} \quad  
z -2 P_N(x) + {4\Lambda^{2N}\over z} = 0.}
This was derived from the constraints,
$$ \int_{A_i} H = N_i = 1  \quad {\rm and} \quad \int_{B_i} H =
\tau_{\rm YM}(\co ) $$
where $\tau_{\rm YM}(\co )$ is the bare Yang-Mills coupling of the original
$\N=2$ theory.
It is clear from the definition of $z$ that ${dz\over z}$ is independent of
$g_{N+1}$ as required.

Finally, substituting \swone\ in \ce\ we get,
$$Z_m =  \oint_{\beta_i}x{dz\over z} - \oint_{\beta_j}x {dz\over z}$$
from which it is possible to identify the Seiberg-Witten differential,
$$ \lambda_{\rm SW} = x {dz\over z} = x{P'_N(x)dx\over \sqrt{P_N^2(x)-4\Lambda^{2N}}}$$
and the mass of the magnetic monopole as $m=|a_{Di} - a_{Dj}|$.

The electric particle is harder to identify in the dual geometry. However,
using the identification between $U(1)^N$ in the original theory and the
$U(1)^N$ in the dual geometry, we can use the charges as a hint to identify
the state. $W_{ij}$-boson are charged under $U(1)_i\times U(1)_j$ with
charges $(1,-1)$. Therefore, it is natural to propose that
the fundamental string stretched between the D5-brane at $x=a_i$ and the
D5-brane at $x=a_j$ corresponds to a fundamental string stretched between a
D3 brane wrapping $S^3_i$ and an anti-D3 brane wrapping $S^3_j$. The
consistency of this argument relies heavily on the fact that there is one
unit of $H_{RR}$ flux through each $S^3$, leading to a fundamental
string charge once a D3 brane is wrapped over it.
As for the BPS mass for electric states,
this should agree with the gauge predictions, because we have already
argued that $a_{Di}$ and $\tau_{ij}$ agree and we have the fundamental
$\N=2$  relation
$${\del a_{D_i}\over \del a_{j}}=\tau_{ij}.$$

\subsec{Generalizations}

A natural generalization of these ideas is to consider the
case of the quiver theories studied in \cfkiv.  We
leave the study of this large class of examples to the reader.
It is quite satisfactory to see this merging of holomorphic techniques
in studying
vacuum structures
for $\N=1$ and $\N=2$ gauge systems via a geometric realization
in string theory and it would be worthwhile studying
more examples of how this works, which we leave to the interested
reader.

\appendix{A}{Proof of $g_{n+1}^2 F_{2n}(x) =W'(x)^2 + f_{n-1}(x)$}

In this section we will review the proof given in \civ\ 
for the reformulation of the superpotential extremization
in the gauge theory setup.
The idea is to formulate the whole problem in terms of a superpotential
with the conditions for massless monopoles imposed as constraints.
Clearly the condition,
\eqn\hha{ P_N(x)^2 -4\Lambda^{2N} = F_{2n}(x)H_{N-n}^2(x) }
is equivalent to,
$$ P_N(p_i) + \epsilon_i 2\Lambda^N  = 0 \quad {\rm and} \quad P'_N(p_i) = 0 $$
for $H_{N-n}(x) = \prod_{i=1}^{N-n}(x-p_i)$ and $\epsilon_i = \pm 1$.

The total superpotential can then be written as,
\eqn\fulli{ W = \sum_{i=1}^n g_r u_r +\sum_{i=1}^{N-n} \left[ L_i
(\left. P_N(x)\right|_{x=p_i}-2\epsilon_i \Lambda^N)+Q_i{\del\over \del
x}\left. P_N(x)\right|_{x=p_i} \right] }
Notice that $l$ is arbitrary now but it will turn out to be $l\ge N-n$. The
$L_i$, $Q_i$, and $p_i$ are treated as Lagrange multipliers.

The variation of \fulli\ with respect to $p_i$ gives
\eqn\pivar{Q_i{\partial ^2 P_N\over \partial x ^2}\bigg| _{x=p_i}=0,}
where we used the $Q_i$ constraint to eliminate the term involving $L_i$.
For generic $g_r$, the RHS of \hha\ has some double roots, but no
triple or higher roots; therefore \pivar\ implies that $\ev{Q_i}=0$.
Since the $\ev{Q_i}=0$, the variation of
\fulli\ with respect to all $u_r$ is
\eqn\wLMur{g_r+\sum _{i=1}^{N-n}\sum _{j=0}^NL_ip_i^{N-j}{\partial s_i\over
\partial u_r},}
with the understanding that the $g_r=0$ for $r>n+1$.  Using that 
$P_N(x) =<{\rm det}(xI-\Phi )>$ and 
$${\rm det}(xI-\Phi ) = \left. x^N {\rm exp}
\left( {\rm tr}\;{\rm log}(I-{1\over x}\Phi)\right) \right|_+ \! =
\left. x^N {\rm exp}\left(-\sum_{n=1}^\infty {u_n\over x^n}\right)\right|_+ 
\! = \left. \sum_{l=0}^{\infty}x^{N-l}s_l \right|_+ $$
where $\left. \sum_{k=-\infty}^{\infty}c_k x^k\right|_+ = 
\sum_{k=0}^{\infty} c_k x^k$,
one can easily show that, ${\del s_j\over \del u_k} = -s_{j-k}$. Therefore,
\wLMur\ becomes
\eqn\wLMurr{g_r=\sum _{i=1}^{N-n}\sum _{j=0}^NL_ip_i^{N-j}s_{j-r}.}
We should also impose the $L_i$ and $Q_i$ constraints in \fulli.  These
equations and \wLMurr\ fix the $\ev{u_r}$, $\ev{L_i}$, $\ev{p_i}$, and
$\ev{Q_i}$ as functions of the $g_r$ and $\Lambda$.  The $\ev{L_i}$
are proportional to the expectation values $\ev{q_i\widetilde q_i}$ of
the $l\geq N-n$ condensed, mutually local, monopoles.

Following a similar argument in \JBYO, we multiply \wLMurr\ by $x^r$
and sum:
\eqn\Wderi{\eqalign{W'(x)&=\sum _{r=1}^Ng_rx^{r-1}
\cr & =\sum _{r=1}^N\sum _{i=1}^{l}\sum_{j=0}^Nx^{r-1}p_i^{N-j}s_{j-r}L_i
\cr &=\sum _{r=-\infty }^N\sum _{i=1}^l\sum_{j=0}^Nx^{r-1}p_i^{N-j}
s_{j-r}L_i-2L\Lambda ^{N}x^{-1}+
{\cal O}(x^{-2})\cr
&=\sum _{i=1}^l\sum_{j=-\infty }^N P_N(x;\ev{u})x^{j-N-1}p_i^{N-j}
L_i-2L\Lambda ^N x^{-1}+{\cal O}(x^{-2})\cr
&=\sum _{i=1}^l{P_N(x;\ev{u})\over x-p_i}L_i-2L\Lambda ^Nx^{-1}
+{\cal O}(x^{-2}).}}
We define $L\equiv \sum _{i=1}^l L_i\epsilon _i$.
Defining, as in \JBYO, the order $l-1$ polynomial $B_{l-1}(x)$ by
\eqn\Bdefn{\sum_{i=1}^l{L_i\over x-p_i}={B_{l-1}(x)\over H_l(x)},}
with $H_l(x)$ the polynomial appearing in \hha,
we thus have
\eqn\wderii{W'(x)+2L\Lambda ^Nx^{-1}=B_{l-1}(x)
\sqrt{F_{2N-2l}(x)+{4\Lambda ^{2N}\over H_l(x)^2}}+{\cal O}(x^{-2}).}
Since the highest order term in $W'(x)$ is $g_{n+1}x^n$,
we see that $B_{l-1}(x)$ should
actually be order $n-N+l$.  This shows that $l\geq N-n$ and, in
particular, for $l=N-n$, $B_{N-n-1}=g_{n+1}$ is a constant.
Squaring \wderii\ gives
\eqn\squarei{g_{n+1}^2F_{2n}=
W'(x)^2+4g_{n+1}L\Lambda ^Nx^{n-1}+{\cal O}(x^{n-2}).}
We have found, $g_{n+1}^2F_{2n}=W'(x)^2+f_{n-1}(x)$.

Notice that after varying with respect to all the Lagrange multipliers and
solving the equations; $<L_i>$, $<Q_i>$, and $<p_i>$ will be functions of $g_i$
and $\Lambda$.

Let us now proof the statement made at the end of section 2. There it was
claimed that $-4 g_{n+1}<S_1+\ldots +S_{n}>$ is equal to the coefficient of the
$x^{n-1}$ term in $f_{n-1}(x)$.

Consider the term in the superpotential \fulli,
$$\sum_{i=1}^{N-n}\left( -2\epsilon_i L_i\right)
 \Lambda^N \equiv -2L \Lambda^N.$$
This tells us that after integrating out $p_i$'s, $Q_i$'s, $u_i$'s, and
$L_i$'s, and $W$ becomes equal to $W_{\rm low}$, then,
$$ {\del W_{\rm low} \over \del {\rm log}\Lambda^{2N}}={\del W \over \del {\rm log}\Lambda^{2N}} = -\Lambda^{N}<L>.$$
{}From \vevs,
$$ {\del W_{\rm low}\over \del {\rm log}\Lambda^{2N}} =<S_1+\ldots +S_{n}>$$
we get,
$$ \sum_{i=1}^{n}<S_i> = -\Lambda^{N}<L>.$$

Finally, using \squarei\ we see that,
$$ f_{n-1}(x) = 4 g_{n+1}L\Lambda^N x^{n-1} + {\cal O}(x^{n-2}).$$
{}From which the statement we wanted to prove follows.

\appendix{B}{Proof of Uniqueness}

We want to understand to what extent our answer for the curves is unique, let us assume
that the following equation holds,
$$ {\tilde W}'(x)^2+{\tilde b}_{n-1}x^{n-1}+\ldots
+{\tilde b}_0  = W'(x)^2+b_{n-1}x^{n-1}+\ldots
+b_0 $$
where $W'(x)=x^n + s_1 x^{n-1}+\ldots +s_{n}$ and ${\tilde W}'(x)=x^n +
{\tilde s}_1 x^{n-1}+\ldots +{\tilde s}_{n}.$

Consider the Riemann surface defined by,
\eqn\hyper{ y^2 =  W'(x)^2+b_{n-1}x^{n-1}+\ldots
+b_0.}
It is not difficult to show that if $C$ is a closed contour around
$x=\infty$ on the upper sheet that does not contain any of the branching
points, then,
\eqn\res{ s_{k} = {1\over 2\pi i}\oint_C x^{k-1-n} y dx }
This can be shown by expanding $y(x)$ 
around $x=\infty$ and reading the residue.  From \res\ we 
conclude that $s_{k} = {\tilde s}_{k}$.

It is also possible to see that,
$${1\over 2} b_{l} = {1\over 2\pi i}\oint_C (x^{l-1}+s_1 x^{l-2}+\ldots
+s_{l-1}) y dx.$$
Therefore, using that $s_{k} = {\tilde s}_{k}$ we conclude that
$b_l={\tilde b}_l$.

We have shown that if the hyperelliptic curve can be written as \hyper\
then the form of the curve is unique.

\appendix{C}{Calculability}

We have shown how the superpotential equations are equivalent to finding a
solution to problem stated in \problem. However,
we have not shown how this can be
used to find the solution.
In this appendix we will show that 
the equations from
our effective superpotential are solvable in a systematic expansion in
$\Lambda$ around the semi-classical regime.

The superpotential \dualsup,
$$W_{\rm eff} = \sum_{i=1}^n N_i \Pi_i + \tym \sum_{i=1}^n S_i$$
is only a function of $S_i$'s. The periods over the non-compact cycles can
be computed in terms of the prepotential of the CY 3-fold $\F = \F
(S_1,\ldots , S_n)$ by $\Pi_i = {\del \F (S_k's)\over \del S_i}$.

The main advantage of the geometric approach is that the prepotential does
not depend on $N_i$'s and once it is found the problem is solved for any
splitting $N=\sum_{i=1}^n N_i$.

The semi-classical approximation in geometric language means that the
deformation of $W'(x)^2 = \Pi_{i=1}^n (x-a_i)^2$ by $W'(x)^2+f_{n-1}(x) =
\Pi_{i=1}^n(x-a_i^+)(x-a_i^-)$ is such that, $|a_i-a_j|\gg |a_k^+-a_k^-|$
for any $\{ i,j,k\}$. The first step is to rewrite the effective one form,
$$\lambda_{\rm eff} =\sqrt{W'(x)^2+f_{n-1}(x)} = \prod_{i=1}^n \sqrt{(x-A_i)^2-\delta_i^2}$$

The compact and non-compact periods $S_j$ and $\Pi_j$ are computed by
changing variables to $y=x-A_j$ and writing,
$$ \lambda_{\rm eff}= \sqrt{y^2-\delta_j^2}\prod_{k\neq j}(y+A_j-A_k)
\prod_{i\neq j}\sqrt{1-\left({\delta_j\over y+A_j-A_i} \right)^2}. $$
Expanding the square roots in the product one gets an infinite power series
in $\delta_i$'s times $\sqrt{y^2-\delta_j^2}$. Integrals of the form,
$$ \int \sqrt{y^2-\delta_j^2}\prod_{k\neq j}(y+A_j -A_k)^{l_k}$$
where $l_k$ are arbitrary integers, can be done in closed form.

The second step is to write the new variables $\{ A_1 ,\ldots , A_n,
\delta_1, \ldots ,\delta_n\}$ in terms of mixed ones $\{a_1,\ldots
,a_n, \delta_1,\ldots ,\delta_n\}$. This is done by equating the $x^{2n},
\ldots ,x^{n}$ coefficients of $W'(x)^2+f_{n-1}(x)$ and of $\prod_{i=1}^n
((x-A_i)^2-\delta_i^2)$. This can be done order by order in $\delta$'s by
solving {\it linear} systems of equations.

Finally, using $S_i$'s as functions of $\{a_1,\ldots
,a_n, \delta_1,\ldots ,\delta_n\}$ one can invert the relations to get
$\delta_k$'s as functions of $\{a_1\ldots ,a_n,S_1\ldots ,S_n
\}$. Substituting this in $\Pi_i = \Pi_i(a_1,\ldots
,a_n, \delta_1,\ldots ,\delta_n)$ one gets $\Pi_i = \Pi_i(a_1\ldots
,a_n,S_1\ldots ,S_n )$. The inversion process can also be done order by
order by solving {\it linear} systems of equations.

In \civ\ this procedure was carried out for $n=2$ with the following
result,
$$
\eqalign{\Pi_1 = & \ldots +S_1 (\log{S_1\over g\Delta } - 1) + 2S_2\log\Delta - 2
(S_1+S_2)\log \co + \cr
& + g(\Delta )^3 \left[{1\over (g\Delta^3)^2}\left( 2S_1^2-10S_1 S_2+5 S_2^2\right)
+
{\cal O}\left({S^3 \over (g\Delta^3)^3}\right) \right]}
$$
where the ellipses stand for terms independent of $S_i$'s, $\Delta =
a_1-a_2$, $g=g_3$ and $\co$ is a large distance cut off. $\Pi_2$ can be
found by replacing all $1$'s by $2$'s and vice versa.

In this case the parameters of the classical superpotential at each order
only enter in an overall coefficient $(g\Delta^3)^{-n}$.

A much more interesting structure can be found for $n=3$. In order to give
the expression of the non-compact periods, a small change in notations has
been introduced. $(1,2,3)$ will be replaced by $(a,b,c)$ and $W'(x) =
(x-a_1)(x-a_2)(x-a_3)$ will be replaced by
$W'(x)=(x-\alpha)(x-\beta)(x-\gamma )$. In terms of the new notation we have,
$$\Pi_a =
\ldots -S_a(1-\log S_a)+(2S_b-S_a)\log(\a-\b)+(2S_c-S_a)\log(\a-\b)
-2(S_a+S_b+S_c)\log\Lambda
$$
$$
+h_{aa}S_a^2+h_{bb} S_b^2+h_{cc} S_c^2+
h_{ab}S_aS_b+h_{ac}S_aS_c + h_{bc}S_bS_c + {\cal O}(S^3)$$
with,
$$\eqalign{ h_{aa}= & {1\over 2(\a-\b)^2(\a-\g)^2}\left( 5+4{(\a-\g)\over (\a
-\b)}+4{(\a-\b)\over (\a-\g)} \right) \cr
h_{bb}= & -{1\over (\b-\a)^2(\b-\g)^2}\left( 2+2{(\g-\b)\over
(\g-\a)}+5{(\b-\g)\over (\b-\a)}\right)\cr
h_{cc}= & -{1\over (\g-\a)^2(\g-\b)^2}\left( 2+2{(\b-\g)\over
(\b-\a)}+5{(\g-\b)\over (\g-\a)}\right)\cr
h_{ab}= & -{2\over (\a-\b)^2(\a-\g)(\b-\g)}\left(-2+5{(\b-\g)\over
(\b-\a)}-2{(\g-\b)\over (\g-\a)}\right)\cr
h_{ac}= & -{2\over (\a-\g)^2(\a-\b)(\g-\b)}\left(2-5{(\g-\b)\over
(\g-\a)}+2{(\b-\g)\over (\b-\a)} \right) \cr
h_{bc}= & {8\over (\a-\b)(\a-\g)(\b-\g)^2}\left( 1-{(\b-\g)\over
(\b-\a)}-{(\g-\b)\over (\g-\a)}\right)} $$
and $\ldots$ represent the classical part $W_{\rm tree}(\a)$ and the diverging
pieces that are $S$-independent.

One can now solve the superpotential equations, ${\del W_{\rm
eff}\over \del S_i}=0$ for a given splitting $N=\sum_{i=1}^n N_i$. This can
be done order by order and gives $<S_i> = <S_i>(a_1,\ldots ,a_n,\Lambda )$.
Using this result one can go back and compute $b_k=b_k(a_1,\ldots
,a_n,\Lambda )$ and $W_{\rm low}=W_{\rm eff}(<S_i's>)$.

\centerline{\bf Acknowledgements}
We would like to thank K. Intriligator and S. Katz
for valuable discussions.  CV would also like
to thank the hospitality of the YITP at Stony Brook.

The research of FC and CV is supported in part by NSF grants PHY-9802709
and DMS-0074329.
\listrefs

\end